\documentclass[OSspectrum,nonblindrev]{informs3}

\DoubleSpacedXI 

\usepackage{endnotes}
\let\footnote=\endnote

\TheoremsNumberedThrough     
\ECRepeatTheorems

\EquationsNumberedThrough    

\usepackage{amsmath}

\begin{document}


\RUNAUTHOR{M. Landete, J.F. Monge and J.L Ruiz}
\RUNTITLE{Determining the best Sharpe ratio portfolio using  a cross-efficiency evaluation}

\TITLE{Sharpe portfolio  using a cross-efficiency evaluation}

\ARTICLEAUTHORS{%
\AUTHOR{Mercedes Landete}
{Centro de Investigaci\'{o}n Operativa, Universidad Miguel Hern\'{a}ndez, Elche (Alicante), Spain, \EMAIL{landete@umh.es}} 
\AUTHOR{Juan F. Monge}
{Centro de Investigaci\'{o}n Operativa, Universidad Miguel Hern\'{a}ndez, Elche (Alicante), Spain, \EMAIL{monge@umh.es}} 
\AUTHOR{Jos\'e L. Ruiz}
{Centro de Investigaci\'{o}n Operativa, Universidad Miguel Hern\'{a}ndez, Elche (Alicante), Spain, \EMAIL{jlruiz@umh.es}} 
} 

\ABSTRACT{%

 The Sharpe ratio is a way to compare the excess returns (over the risk free asset) of portfolios for each unit of volatility that is generated by a portfolio. In this paper we introduce a robust Sharpe ratio portfolio under the assumption that the risk free asset is unknown. We propose a robust portfolio that maximizes the Sharpe ratio when the risk free asset is unknown, but is within a given interval. To compute the best Sharpe ratio portfolio all the Sharpe ratios for any risk free asset are considered and compared by using the so-called cross-efficiency evaluation. An explicit expression of the Cross-Eficiency Sharpe ratio portfolio is presented when short selling is allowed.
}
\KEYWORDS{finance, portfolio, minimum-variance porfolio, cross-efficiency}

\maketitle

%


{\bf Funding:} The authors are grateful to the Spanish Ministry of Economy and Competitiveness for supporting this research through grant MTM2013-43903-P.

\section{Introduction}

In 1952 Harry Markowitz made the first contribution to portfolio optimization. In the literature on asset location, there has been significant progress since the seminal work by Markowitz in 1952, \cite{markowitz52}, who introduced the optimal way of selecting assets when the investor only has information about the expected return and variance for each asset in addition to the correlation between them.

In 1990, Harry Markowitz, Merton Miller and William Sharpe won the Nobel Prize in Economics for their portfolio optimization theory.

The optimal portfolio obtained by the Markowitz model usually shows high long-term volatility. This feature has motivated a body of research oriented to control the present error in the Markowitz model. Since the variance of the portfolio cannot be considered as an adequate measure of risk, a number of alternative measures have been proposed in the literature in an attempt to quantify the portfolio variance more appropriately (see \cite{markowitz59,Jin2006,Nawrocki} among others). Another way to control the risk in the optimization model is based on setting a minimum threshold for the expected return. Following that approach, several models which incorporate risk measures such as ``safety measure'', ``value at risk'', ``conditional value at risk'', etc., have been proposed in order to control the volatility of the solution. See \cite{Artzener99,Krokhmal2002} and references therein.

The incorporation of new restrictions to the problem is also a tool that has been used both to prevent the risk and to incorporate the knowledge of the analyst in search of the best solution. New models have emerged in the last years, which include linear programming models, integer optimization models and stochastic programming models (see \cite{mansini2014}  among others).

Another important feature of the Markowitz model is its myopia about the future scenario of potential returns that will happen. For this reason, producing accurate forecasts in portfolio optimization is of outmost importance. In this sense, forecasting models, factor models in covariance matrix and return estimation, bayesian approaches, or uncertainty estimates (see \cite{BenTal1999} and references therein) are helpful. The need to improve predictions and consider the present uncertainty in the Markowitz model has motivated the development of what is collectively known as ``robust optimization'' techniques. Robust methods in mathematical programming were introduced by [3] and after studied in a portfolio context by \cite{goldfarb2003} among others.

There exist several methods in the literature aimed at improving the performance of Markowitz’s model, but none of these methods can be considered better than the others. To the author’s knowledge, a systematic comparison of the approaches discussed above has not yet been published. However, in \cite{demiguel2009_2} 14 different models are compared on the basis of a number of datasets with different quality measures. The results obtained show that {\em ``none of the sophisticated models consistently beat the na\"\i{}ve 1/N benchmark''}.

Our objective in this paper is to determine the best tangent portfolio, when the free risk rate asset is unknown or the information on this parameter is not deterministic for a long time period.  The goal is to find a robust portfolio in the sense of a tangent portfolio better than other tangent portfolios compared withit. To achieve that goal, we use some techniques based on Data Envelopment Analysis (DEA), which provides an analysis of the relative efficiency of the units involved. In the context of portfolio optimization, several authors have used such DEA techniques, specifically the cross-efficiency evaluation (like us here), yet with a different purpose (see, for example, \cite{lim2014}).

In the next section we present a brief description of the original Markowitz and Sharpe ratio models for portfolio optimization, and discuss some of the features related to the solutions and the efficient frontier that will be needed for the remainder of the paper. In Section 3 we propose an approach to portfolio optimization based on the Cross-Efficiency evaluation. In Section 4 we compare our approach with other classical solutions through the study of two pilot cases. And in the last section  we offer a conclusion.

\section{Overview}\label{section1}

In this section, we present a brief description of the Sharpe ratio for asset allocation. The portfolio optimization problem seeks a best allocation of investment among a set of assets. The model introduced by Markowitz provides a portfolio selection as a return-risk bicriteria tradeoff where the variance is minimized under a specified expected return. The mean-variance portfolio optimization model can be formulated as follows:

\begin{align}
\min  \qquad   \sigma_P^2 &=\frac{1}{2}w^T \Sigma w  \label{fo.m1}
&& \\
s.t.   \qquad   w^{ T} \mu &= \rho  \label{r1.m1}\\
             w^{ T} 1_n &=1  \label{r2.m1}
\end{align}

The objective function (\ref{fo.m1}),  $\sigma_P^2$,   gives the variance of the return $w^T \mu$, where $\Sigma$ denotes the $n\times n$ variance-covariance matrix of $n-$vector of returns $\mu$, and $w$ is the $n-$vector of portfolio weights invested in each asset. Constraint (\ref{r1.m1}) requires that the total return is equal to the minimum rate $\rho$ of return the investor wants. The last constraint (\ref{r2.m1}) forces to invest all the money. We denote by     $1_n$ the $n-$dimensionall vector of ones. Note that the weight vector $w$ is not required to be non-negative as we want to allow short selling, whose weight of vector $w$ is less than 0.

\begin{figure}[t]
\begin{center}
					\includegraphics[width=.7\textwidth]{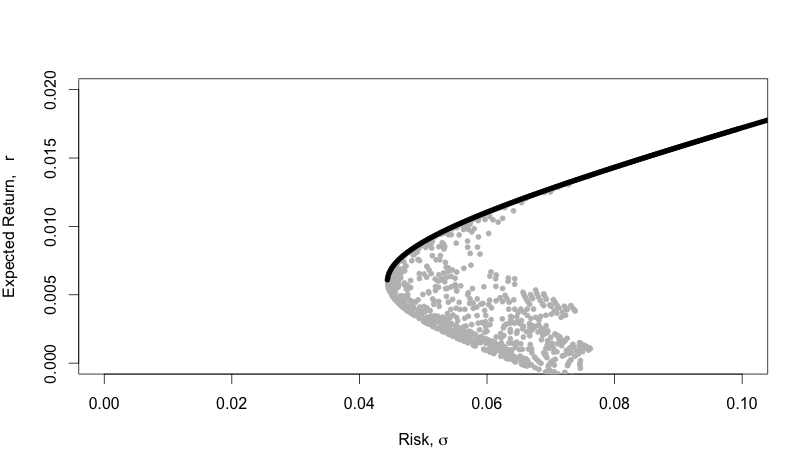}
				\caption{Efficient frontier and cloud of possible portfolios.}
				\label{fig1}
				\end{center}
\end{figure}

This model uses the relationship between mean returns and variance of the returns to find a minimum variance point in the feasible region. This minimum variance is a point on the $efficient$ $frontier$, $W_\rho$. The efficient frontier is the curve that shows all efficient portfolios in a risk-return framework, see Figure \ref{fig1}.

\subsection{Global minimum variance portfolio}

The {\bf Global Minimum Variance}  (GMV)  portfolio from the Efficient Frontier ($W_\rho$) is obtained without imposing the expected-return constraint (\ref{r1.m1}). The portfolio weights, ($w_{GMV}^*$), expected return ($r_{GMV}^*$) and variance ($\sigma_{GMV}^{*2}$) are given by 
\begin{eqnarray}
w^*_{GMV}=\frac{\Sigma^{-1} 1_n}{1_n^T \Sigma^{-1}1_n}, \qquad 
r^*_{GMV}=\frac{1_n^T\Sigma^{-1}\mu }{1_n^T \Sigma^{-1}1_n}  \qquad \text{and}\qquad 
\sigma^{*2}_{GMV}=\frac{1}{1_n^T \Sigma^{-1}1_n} 
\label{sol_GMV}
\end{eqnarray}

\begin{figure}[htb]
\begin{center}
					\includegraphics[width=.7\textwidth]{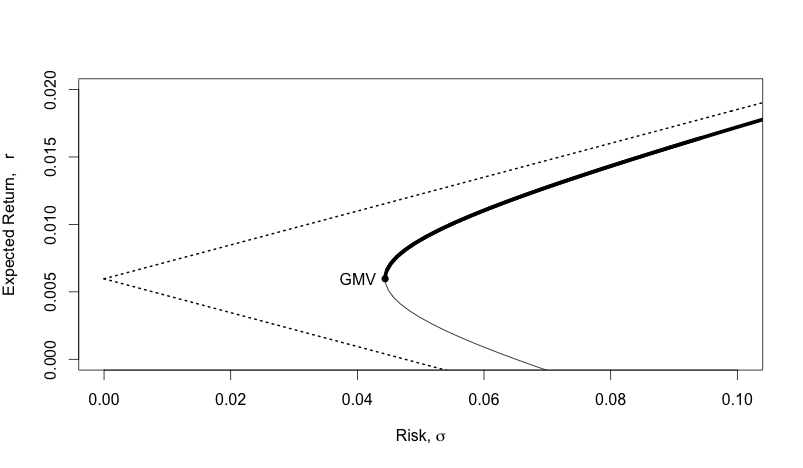}
								\caption{Hyperbola and the assymptotes for mean-variance efficient portfolios.}
				\label{fig2}
\end{center}
\end{figure}

The hyperbola of the feasible portfolios is enclosed by the the asymptotes  $r=c/b \pm \sqrt{(ab-c^2)/b}\, \sigma$ with 

\begin{align}
a=\mu^T \Sigma^{-1} \mu, \quad b=1_n^T \Sigma^{-1} 1_n \quad \text{and} \quad c=1_n^T \Sigma^{-1} \mu.
\label{abc}
 \end{align}

The expected return of the global minimum variance portfolio, $r_{GMV}$, is the apex of the hyperbola.  Figure \ref{fig2} represents the hyperbola for the feasible portfolios, the efficient frontier, the global minimum variance portfolio (GMV) and the asymptotes.

\subsection{Sharpe ratio}

The {\bf Tangent Portfolio} (TP) is the portfolio where the line through the origin is tangent to the efficient frontier $W_\rho$. This portfolio represents the portfolio with maximum ratio mean/variance.

\begin{align}
 w^*_{TP} =arg \max_{w } \frac{w^T \mu}{\sqrt{w^T \Sigma w}} \quad  \label{TP.m}
 s.t. \quad w^T1_n=1
 \end{align}

Another studied portfolio is obtained by maximizing the same ratio when a risk free asset, $r_f$, is considered. This portfolio is called the {\bf Maximum Sharpe Ratio} (MSR) portfolio. The Sharpe ratio is the expected excess returns (over the risk free asset) per unit of risk. Therefore, the Maximum Sharpe Ratio (MSR) portfolio is the solution to the model:

\begin{align}
w^*_{MSR}     =arg \max_{w } \frac{w^T (\mu-r_f)}{\sqrt{w^T \Sigma w}}  \label{MSR.m}
\quad  s.t. \quad w^T1_n=1
\end{align}

where $r_f$ denotes the risk free asset. The allocation $w^*_{MSR}$ is known as market portfolio, $M$. If the risk free rate is $r_f = 0$, the market portfolio is identical to the tangent portfolio solution of problem (\ref{TP.m}). 

{\bf Capital Market Theory} asks about the relationship between expected returns and risk for portfolios and free-risky securities.

\begin{figure}[h]
\begin{center}
					\includegraphics[width=.9\textwidth]{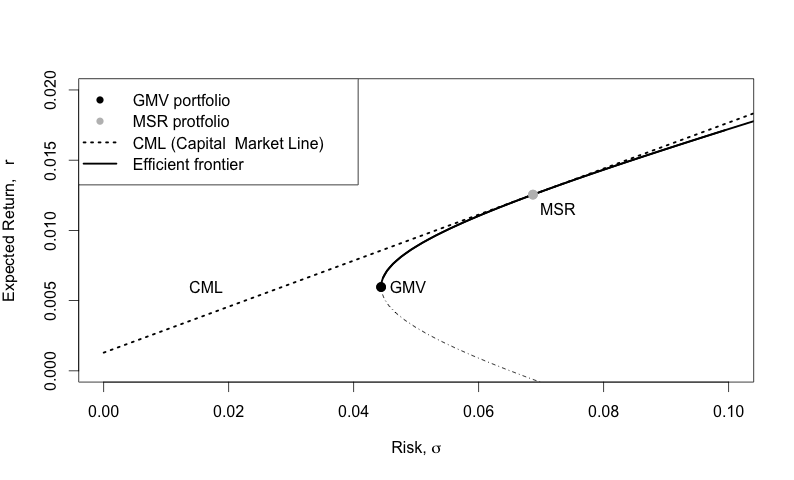}
				\caption{Efficient frontier obtained from four assets. The Global Minimum Variance (GMV) is the portfolio with less risk. The Maximum Sharpe Ratio (MSR) is the tangent portfolio located in the Efficient frontier in the presence of a risk-free asset. The combination of the risk-free asset and the tangency portfolio (MSR) generates the Capital Market Line (CML). CML is the set of non-dominated portfolios when a risk-free asset is present.}
								\label{fig3}
\end{center}
				
\end{figure}

The solution to (\ref{MSR.m}) includes only risky assets. This solution is known as the Market Portfolio ($M$). A line from the risk-free interest rate through the Market portfolio ($M$) is known as the {\bf Capital Market Line} ($CML$). All the efficient portfolios must lie along this line,
 \[ CML: \quad E(r)=r_f + \frac{r_M-r_f}{\sigma_M}\sigma\]

where $E(r)$ is the expected portfolio return, $r_f$ the risk-free rate of interest, and $r_M$, $\sigma_M$, respectively, the return and risk of the Market portfolio $M$. All the portfolios on the $CML$ have the same Sharpe ratio. See figure \ref{fig3}.

The $CML$ summarizes a simple linear relationship between the expected return and the risk of efficient portfolios. Sharpe assumed that the total funds were divided between the Market portfolio ($M$) and security $f$ . The inversion is fully invested here.

\[ w_{M}+w_{f}=1\]
The expected return of the portfolio is 
\[ E(r_p)=w_f r_f + w_M r_M\]

In order to calculate the optimal $MSR$ portfolio of (\ref{MSR.m}) we have to maximize (\ref{MSR.m}) subject to $w^T1_n=1$. In \cite{chapados2011} we can see how to derive the following expression for the solution of this problem:

\begin{align}
 w^*_{MSR} = \frac{\Sigma^{-1} (\mu -r_f)}{1_n^T \Sigma^{-1} (\mu-r_f)}
 \end{align}

The risk $\sigma^*$ and the expected excess returns $r^∗$ for the optimal solution to the maximization Sharpe ratio problem with free risk $r_f$ is:

\begin{align}
r_{MSR}^*&=w_{MSR}^{*T } \, \mu =  \frac{(\mu-r_f)^T \Sigma^{-1} \mu }{1_n^T \Sigma^{-1} (\mu-r_f)} \label{sol1_MSR}&\\
\sigma^*_{MSR} & =\sqrt{w_{MSR}^{*T } \, \Sigma  \, w_{MSR}^*} =\frac{ \sqrt{ (\mu-r_f)^T \Sigma^{-1} (\mu -r_f)}}{1_n^T \Sigma^{-1} (\mu-r_f)}  \label{sol2_MSR}&
 \end{align}

Tangent portfolios are portfolios usually designed for long-term investors. Most investors tend to take on too much risk in good times, and then sell out in bad times. Tangent Portfolios are designed to let investors  do well enough in both good and bad times. This lets us reap the long-term benefits of investing in stocks and bonds with a simple, low-maintenance solution.

\begin{figure}[h]
\begin{center}
					\includegraphics[width=.7\textwidth]{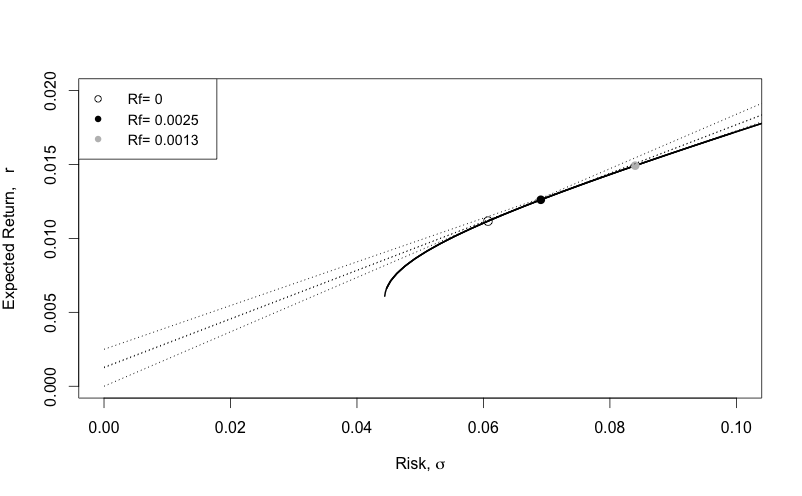}
				\caption{Different CML lines  when the  risk-free asset  is 0, 0.0013 and 0.0025, and their  maximum Sharpe ratio portfolios.}
									\label{fig4_1}
\end{center}
\end{figure}

\begin{figure}[h]
\begin{center}
					\includegraphics[width=.7\textwidth]{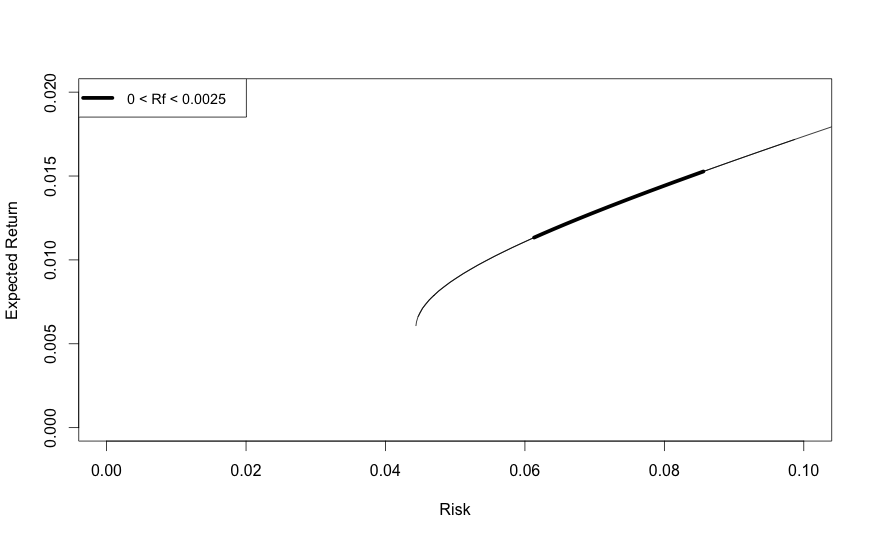}
				\caption{$W_{\rho}^{r_f}$ set when  the  risk-free asset  is in the interval [0, 0.0025].}
									\label{fig4_2}
\end{center}
\end{figure}

Denote by $W_{\rho}^{r_f}$ the subset of efficient portfolios, $W_\rho^{r_f} \subset W_{\rho} $, formed for maximum  Sharpe ratio portfolios, i.e., tangent portfolios obtained for some value of $r_f$. Note that, all the tangent portfolios in $W_{\rho}^{r_f}$ are obtained varying $r_f$ from 0 to the hyperbola apex $r_{GMV}^*$, i.e., $r_f\in [0,r_{GMV}^*]$. See figures \ref{fig4_1} and \ref{fig4_2}.

 \section{Portfolio selection based on a cross-efficiency evaluation}
 
In this section we propose an approach to make the selection of a portfolio within the set $W_{\rho}^{r_f}$. This approach is inspired by the so-called Data Envelopment Analysis (DEA) and cross-efficiency evaluation methodologies. DEA, as introduced in Charnes et al. (1978) evaluates the relative efficiency of decision making units ($DMUs$) involved in production processes. For each $DMU_0$, it provides an efficiency score in the form of a weighted sum of outputs to a weighted sum of inputs. DEA models allow the $DMUs$ total freedom in the choice of the input and output weights. This means that each $DMU_0$ chooses the weights that show it in its best possible light, with the only condition that the efficiency ratios calculated for the other $DMUs$ with those weights are lower than or equal to a given quantity, usually set at 1. Thus, $DMU_0$ is rated as efficient if its efficiency score equals 1. Otherwise, it is inefficient, and the lower the efficiency score, the larger its inefficiency. DEA has been successfully used in many real applications to  analyze efficiency in areas like banking, healthcare, education or agriculture.

 Inspired by DEA, we propose to solve the following model for each portfolio $(\sigma_0,r_o)$  in  $W_{\rho}^{r_f}$ 
 
 \begin{equation}
 \begin{tabular}{ll} 
 & \\  
Maximize &  $ \frac{ \displaystyle u\, (r_0 -u_0)}{\displaystyle v\, \sigma_0} $  \\ 
 & \\
Subjet to.    &   $\frac{\displaystyle u\, (r-u_0) }{\displaystyle v\, \sigma}  \leq 1$ \qquad $\forall (\sigma,r) \in W_{\rho}^{r_f}$ \\
                                          &    $u,v   \geq 0$ \\
                                          & $u_0 \geq 0$ \\   
\end{tabular}
\label{BCC_P.m}
\end{equation}

   
 In (\ref{BCC_P.m}), the portfolios in $W_{\rho}^{r_f}$ would be playing the role of the DMUs, which in this case have one single input (risk, $\sigma$) and one single output (return, $r$). It should be noted that, unlike the problem we address here, in standard DEA we have a finite number of DMUs. Obviously, the optimal value of (\ref{BCC_P.m})  when solved in the evaluation of each portfolio in $W_{\rho}^{r_f}$, $(\sigma_0,r_o)$, equals 1, because there exist non-negative weights  $u^*$, $v^*$  and $v_0^* $    such that   $u^*(r_0-u^*_0)/   v^* \sigma_0 =1$, and  $u^*(r-u^*_0)/   v^* \sigma \leq 1$   for the rest of portfolios $(\sigma,r)\in W_{\rho}^{r_f}$. These weights are actually the coefficients of the tangent hyperplane to the curve (efficient frontier) $W_{\rho}$    at $(\sigma_0,r_0)$.

 Denote in general by $(\sigma^*_{MSR_i},r^*_{MSR_i})\in W_{\rho}^{r_f}$  the MSR portfolio obtained when the risk-free rate $r_f$ is $r_f^i$.    This portfolio maximizes the Sharpe ratio (\ref{MSR.m}). Therefore, the optimal solution of (\ref{BCC_P.m}) when $(\sigma^*_{MSR_i},r^*_{MSR_i})$  is evaluated is

 \begin{equation}
u_i^*=\frac{1}{r^*_{MSR_i} - r_f^i }, \quad v_i^*= \frac{1}{\sigma_{MSR_i}^*} \quad \text{and} \quad u^*_{i0}=r_f^i \label{sol_efic_problem}
\end{equation}

As said before, its optimal value equals 1. Nevertheless, these optimal solutions for the weights allow us to define the cross-efficiency score of a given portfolio obtained with the weights of the others.
 
   \begin{definition}
 Let $(u_j^*,v_j^*,u_{j0}^*)$ be an  optimal solution of (\ref{BCC_P.m}) for portfolio $j:=(\sigma^*_{MSR_j},r^*_{MSR_j})$. The cross-efficiency of a given portfolio  $i:=((\sigma^*_{MSR_i},r^*_{MSR_i})$  obtained with the weights of portfolio $j$ is defined as follows
\begin{equation}
 Ef_i(r_f^j)=\frac{u^*_j(r^*_{MSR_i}-u^*_{j0})}{v^*_j\sigma^*_{MSR_i}} \label{CE.m}
 \end{equation}
 \end{definition}

We can see that (\ref{CE.m}) provides an evaluation of the efficiency of portfolio $i$ from the perspective of portfolio $j$.
	
The following proposition holds

\begin{proposition}
\begin{equation}
Ef_i(r_f^j)=\frac{(r^*_{MSR_i}-r_f^j)/\sigma^*_{MSR_i}}{(r^*_{MSR_j}-r_f^j)/\sigma^*_{MSR_j}}
\label{E.m}
\end{equation}
\end{proposition}
\proof{Proof of Proposition 1.} 
 \begin{equation}
    Ef_i(r_f^j)=\frac{u^*_j(r^*_{MSR_i}-u_{j0})}{v^*_j\sigma^*_{MSR_i}}=\frac{\sigma^*_{MSR_j}(r^*_{MSR_i}-r_f^j)}{\sigma^*_{MSR_i}(r^*_{MSR_j}-r_f^j)}=\frac{(r^*_{MSR_i}-r_f^j)/\sigma^*_{MSR_i}}{(r^*_{MSR_j}-r_f^j)/\sigma^*_{MSR_j}} \nonumber
\label{E2.m}
\end{equation} \Halmos
\endproof

 
 (\ref{E.m}) provides a different interpretation of the cross-efficiency scores. Specifically, $Ef_i(r_f^j)$ represents the ratio between the excess return by risk of portfolio $i$ with respect to portfolio j when the risky-free asset is $r_j$ , that is, how bad the Sharpe ratio of portfolio $i$ is compared to the optimal Sharpe ratio of portfolio $j$.
 
Cross-efficiency evaluation (Sexton et al., 1986, and Doyle and Green, 1994) arose as an extension of DEA aimed at ranking DMUs. DEA provides a self-evaluation of DMUs by using input and output weights that are unit-specific, and this makes impossible to derive an ordering. In addition, it is also claimed that cross-efficiency evaluation may help improve discrimination, which is actually the problem we address in the present paper. DEA often rates many DMUs as efficient as a result of the previously mentioned total weight flexibility: DMUs frequently put the weight on a few inputs/outputs and ignore the variables with poor performance by attaching them a zero weight. The basic idea of cross-efficiency evaluation is to assess each unit with the weights of all DMUs instead of with its own weights only. Specifically, the cross-efficiency score of a given unit is usually calculated as the average of its cross-efficiencies obtained with the weights profiles provided by all DMUs. Thus, each unit is evaluated with reference to the range of weights chosen by all DMUs, which provides a peer-evaluation of the unit under assessment, as opposed to the conventional DEA self-evaluation. In particular, this makes possible a ranking of the DMUs based on the resulting cross-efficiency scores. Cross-efficiency evaluation has also been widely used to address real world problems, in particular to deal with issues related to portfolios (see \cite{lim2014,Galagedera2013,Leivo2012}). Next, we adapt the idea of the standard cross-efficiency evaluation to the problem of portfolio
selection we address here. In order to do so, we first define the cross-efficiency score of a given portfolio, which is the measure that will be used for the selection of portfolios among those in $W_{\rho}^{r_f}$.

 \subsection{Cross-efficiency Sharpe ratio portfolio}

 In this section we present the average cross-eficiency measure for any portfolio $(\sigma,r)\in W_{\rho}$ and obtain an expression for the Maximum Cross-Efficiency Sharpe Ratio portfolio (MCESR).

 \begin{definition}
 Let $r_f$ be the risk-free rate, which satisfies  $r_f\in [r_{\min},r_{\max}]$, then the average cross-efficiency score ($CE_i$) of portfolio i,  $i=(\sigma^*_{MSR_i}, r^*_{MSR_i})\in W_{\rho}$ with $r_i \in [r_{\min},r_{\max}]$, is given by:
 \begin{equation}
CE_i=\frac{1}{r_{\max}-r_{\min}}\int_{r_{\min}}^{r_{\max}}Ef_i(r_f)  \, \text{d} r_{f}
\label{CE.f}
\end{equation}
  \end{definition}

 Note that the expression (\ref{CE.f}) is a natural extension of the cross-efficiency evaluation in DEA for a continuous frontier. Using  the expression  of $Ef_i(r_f^j) $, see equation (\ref{E.m}), the cross efficiency $CE_i$ can be written as:
 
  \begin{equation}
CE_i=\frac{r^*_{MSR_i}}{\sigma^*_{MSR_i}} I_1 - \frac{1}{\sigma^*_{MSR_i}} I_2
\label{CE2.f}
\end{equation}
where,

 \begin{align}
&I_1=\frac{1}{r_{\max}-r_{\min}} \int_{r_{\min}}^{r_{\max}}\frac{\sigma^*_{MSR_f}}{r^*_{MSR_f}-r_f} \, \text{d}r_{f} \label{I1.f} \\
&I_1=\frac{1}{r_{\max}-r_{\min}} \int_{r_{\min}}^{r_{\max}}\frac{\sigma^*_{MSR_f} \,\,  r_f}{r^*_{MSR_f}-r_f}  \, \text{d}r_{f} \label{I2.m}
\end{align}

 \begin{proposition} The efficient portfolio $i=(\sigma^*_{MSR_i}, r^*_{MSR_i})$ that maximizes the cross-efficiency $CE_i$, in the interval  $[r_{1},r_{2}]$, is reached when 
  \begin{equation}
r_i^* =  r_{GMV}^*+\sigma_{GMV}^* \frac{  \displaystyle \frac{ \displaystyle r^*_{MSR_{2}} - r_{2}}{\sigma^*_{MSR_{2}}}  - \frac{r^*_{MSR_{1}} - r_{1}}{\sigma^*_{MSR_{1}}}  }{ \ln \left ( \displaystyle \frac{ 
\displaystyle \frac{   \displaystyle r^*_{MSR_{2}} - r_{2}}{ \displaystyle \sigma^*_{MSR_{2}}}  - \frac{ r^*_{GMV} - r_{2}}{\sigma^*_{GMV}  }}{ 
\displaystyle    \frac{ \displaystyle r^*_{MSR_{1}} - r_{1}}{\sigma^*_{MSR_{1}}}  -\frac{ \displaystyle r^*_{GMV} - r_{{1}}}{\sigma^*_{GMV} }      } \right )}
\label{r.f}
\end{equation}
 \end{proposition}
\proof{Proof of Proposition 2.}  See the appendix. \Halmos
\endproof
 
\begin{figure}[h]
\begin{center}
					\includegraphics[width=.7\textwidth]{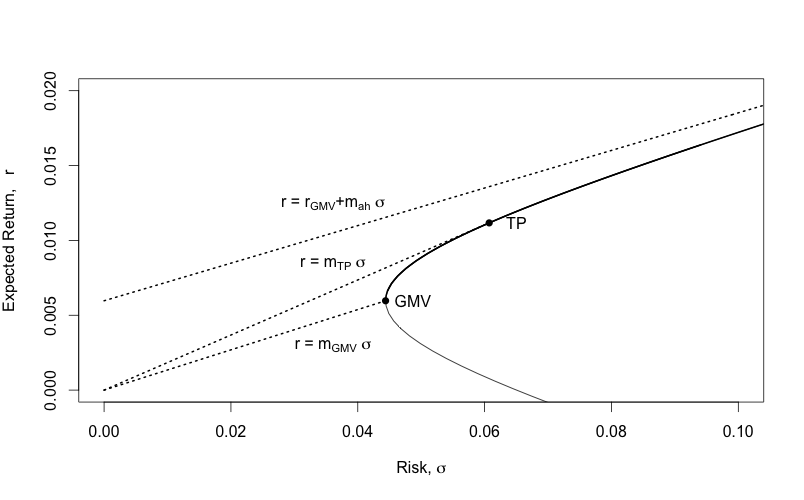}
									\caption{Linear lines for the global minimum variance and tangent portfolios; and for  the asymptote of the hyperbola.}
			\label{fig5}
	\end{center}
\end{figure}

 \begin{corollary} The maximum cross-efficiency (MCESR) portfolio  in the interval $[0,r^*_{GMV}]$ is reached when
   \begin{equation}
r_i^* =  r_{GMV}^*+\sigma_{GMV}^* \frac{  \displaystyle \frac{ \displaystyle r^*_{MSR_{2}} - r^*_{GMV}}{\sigma^*_{MSR_{2}}}  - \frac{r^*_{TP} }{\sigma^*_{TP}}  }{ \ln \left ( \displaystyle  
\displaystyle \frac{   \displaystyle r^*_{MSR_{2}} - r^*_{GMV}}{ \displaystyle \sigma^*_{MSR_{2}}}        \right )     -
 \ln \left ( \displaystyle    \frac{ \displaystyle r^*_{TP} }{\sigma^*_{TP}}  -\frac{ \displaystyle r^*_{GMV}  }{\sigma^*_{GMV} }       \right )  
} 
\label{r2.f}
\end{equation}
and, we can write the above expression as:
   \begin{equation}
r_i^* =  r^*_{GMV}\left ( 1- 
\frac{ \displaystyle   \frac{m_{ah}}{m_{GMV}} - \frac{m_{TP}}{m_{GMV}}   }
{ \ln \left (  \displaystyle \frac{m_{TP}}{m_{ah}}  - \frac{m_{GMV}}{m_{ah}}  \right )     }
\right ), \quad \text{where}
\label{r3.f}
\end{equation}
 \end{corollary}
 $m_{ah} =(r_{MSR_2}^* - r_{GMV}^*)/\sigma^*_{MSR_2}$ is the slope of the asymptote of $W_\rho$, $m_{TP} =r_{TP}^* /\sigma^*_{MSR_2}$ is the slope of the CML line when $r_f=0$, i.e., the slope of the linear line from the origin to the tangent portfolio, and, $m_{GMV} =r_{GMV}^*/\sigma^*_{GMV}$  is the slope of the linear line from the origin to the global minimum variance portfolio (GMV), see figure \ref{fig5}.
\proof{Proof of corollary 1.}   It follows from proposition   2.  \Halmos
\endproof

\begin{proposition} There  exists a Pythagorean  relationship  between the slopes of the Tangent  and Global Minimum  portfolios and the slope of the asymptote of $W_{\rho}$. 
 \begin{equation}
m_{TP}^2= m_{ah}^2+m_{GMV}^2 
\end{equation}
\end{proposition}
\proof{Proof of proposition 3.} See the appendix.
 \Halmos
\endproof

   \begin{corollary} The maximum cross-efficiency (MCESR) portfolio  in $[0,r^*_{GMV}]$ depends only  on Minimal Global Variance and Tangent portfolios. 
   \begin{equation}
r_i^* =  r^*_{GMV}\left ( 1- 
\frac{\displaystyle   \sqrt{\frac{r^*_{TP}}{r^*_{GMV} }}-   \sqrt{\frac{r^*_{TP}}{r^*_{GMV} }-1}     }
{ \ln \left ( \displaystyle  \sqrt{\frac{r^*_{TP}}{r^*_{GMV} } -1}  \right )   - \ln \left (\displaystyle  \sqrt{\frac{r^*_{TP}}{r^*_{GMV} }} -1 \right )   }
\right )
\label{r4.f}
\end{equation}
   \end{corollary}

\proof{Proof of corollary 2.} See the appendix.
 \Halmos
\endproof

 \subsection{No Short-Sales Constraint}
 
 This constraint corresponds to the requirement that all asset weights be non-negative. If no short selling is allowed, then we need to add the non negativity of each weight in vector $w$ to the maximization  Sharpe ratio  problem (\ref{MSR.m}), 
 
 \begin{align}
w^*_{MSR}     = & \max_{w } \frac{w^T (\mu-r_f)}{\sqrt{w^T \Sigma w}}  \label{MSR_NSS.m} \\
&\quad  s.t.  \quad w^T1_n=1  \nonumber \\ 
& \qquad  \qquad \,\,\,\,\, w \geq 0  \nonumber
\end{align}

Markowiz's original formulation (\ref{fo.m1}-\ref{r2.m1}) included those constraints as an integral part of the portfolio optimization method. Note that the inclusion of these non negativity constraints makes impossible to derive an analytical solution for the portfolio optimization problem (\ref{MSR_NSS.m}). Model (\ref{MSR_NSS.m}) is not a convex problem, so it is not easy to solve it. In \cite{Ttnc03optimizationin}, R.H. T\"aut\"aunc\"au present a convex quadratic programming problem equivalent to (\ref{MSR_NSS.m}). This new formulation of the problem considers a higher dimensional space where the quadratic problem is convex when  applying the  lifting technique that follows. 

It is easy to derive the equivalent problem of  (\ref{MSR_NSS.m}) as  

 \begin{align}
 \min \quad  & x^T \Sigma x  \label{MSR_NSS2.m} \\
  s.t. \quad &  x^T (\mu-r_f) =1 \nonumber \\ 
 \qquad &  \qquad\qquad \, x \geq 0  \nonumber
\end{align}
 where the weight vector $w$ of $(\ref{MSR_NSS.m})$ is given by  
 \[ w= \frac{x}{x^T1_n}.  \]
 Note that problem (\ref{MSR_NSS2.m}) can be solved by using the well-known techniques for convex quadratic programming problems. 
 
 Although it is not possible to find a closed  expression for the the Maximum Cross-Efficiency (MCESR) portfolio, model (\ref{MSR_NSS2.m}) allows us to obtain an optimal portfolio, for the maximization Sharpe ratio problem,  and for different values of the risky-free asset.  We propose the following procedure to  obtain an approximation to the Maximum Cross-Efficiency (MCESR) portfolio when no short-sales constrains are present.
 
 \begin{enumerate}
 \item Dividing the interval $[r_{\min}, r_{\max}]$ into $n$ equal parts.
 \item For (i=1 to n+1), solving  (\ref{MSR_NSS2.m}) with $r_f=  r_{\min}*(n-i+1)/n  +  r_{\max}*(i-1)/n$, and obtaining the the efficient portfolio $i=(\sigma_{MSR_i}, r_{MSR_i})$, $\forall i=1$ to $n$.
 \item Computing the solution  $(u^*_i,v^*_i,u^*_{i0})$ of problem (\ref{BCC_P.m}) through expressions  (\ref{sol_efic_problem}). Note that is not necessary to solve the problem  (\ref{BCC_P.m}), the solution of the problem  is the tangent hyperplane to efficient curve $W_{\rho}$. 
 \item For (i=1 to n+1), calculating  the cross-efficiency ($CE_i$) of portfolio $i$  as the mean of the efficiency score of portfolio $i$ by using the optimal weights of the remaining  portfolios in the interval, i.e., 
\begin{equation}
CE_i= \frac{1}{n+1} \sum_{j=1}^{n+1}\frac{(r^*_{MSR_i}-r_f^j)/\sigma^*_{MSR_i}}{(r^*_{MSR_j}-r_f^j)/\sigma^*_{MSR_j}}
\label{cross_effi_NSS}
\end{equation}
\item  Obtaining the efficient portfolio $i$ that maximizes the cross-efficiency $CE_i$. 
 \end{enumerate}

\section{Numerical example}

We carried out two computational studies in order to illustrate the proposed approach. In the first one, we evaluate the goodness of the maximum cross-efficiency portfolio (MCESR) and draw some conclusions. The second part of the study allows us to compare the (MCESR) allocation depending on whether short-sales are allowed or not.

The whole computational study was conducted on a MAC-OSX with a 2.5GHz Intel Core i5 and 4GB of RAM. We used the R-Studio, v0.97.551 with the library {\em stockPortfolio}, \cite{stockportfolio}. In our computational study the required computational time did not exceed a few seconds; for this reason the times have not been reported.

\subsection{Case 1. EUROSTOCK}

In this section we compare the performance of our Maximum Cross-Efficiency Sharpe ratio (MCESR) portfolio allocation with different Sharpe ratio allocations on a small example with real data. The set of assets that were chosen are listed in Table 1, and these were obtained from EUROSTOCK50. We selected the six Spanish assets in EUROSTOCK50.

\begin{figure}[htp]
\begin{center}
					\includegraphics[width=.75\textwidth]{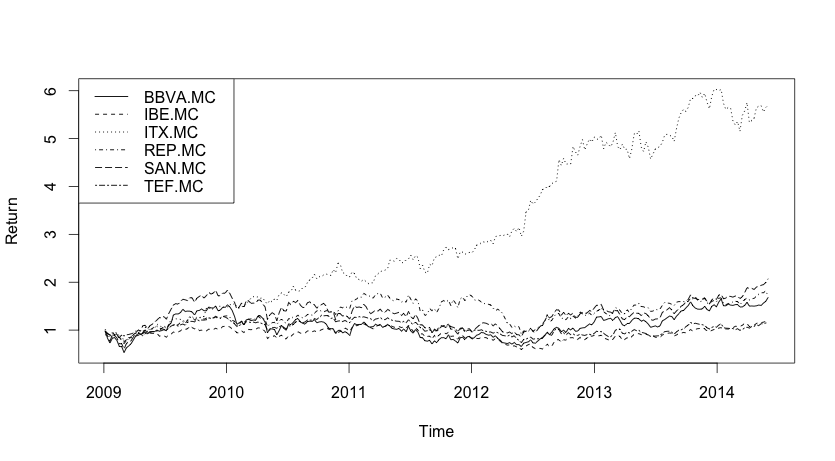}
					\caption{Case 1. Returns for the total period,  from 2009-01-01 to 2014-06-06.}
			\label{fig_s3}
	\end{center}
\begin{center}
					\includegraphics[width=.75\textwidth]{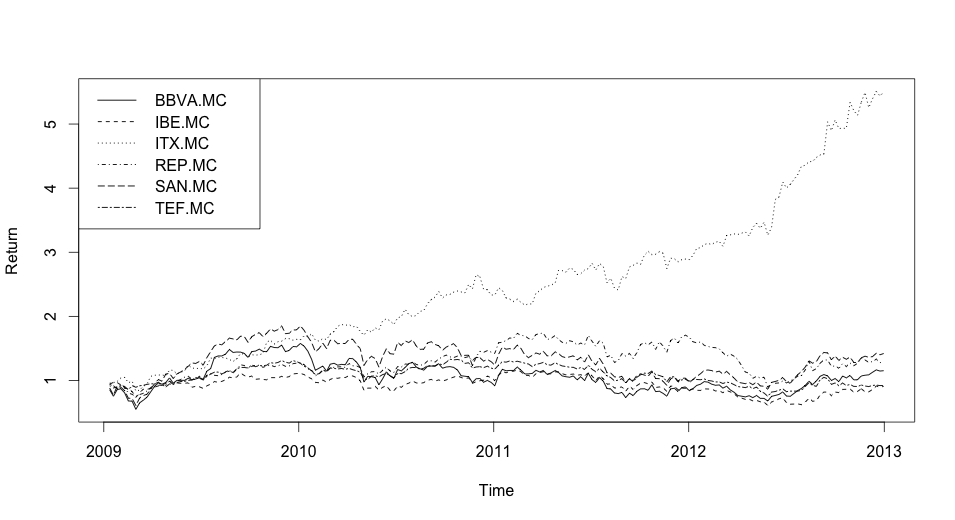}
					\caption{Case 1. Returns for in-sample period,   from 2009-01-01 to 2012-12-31.}
			\label{fig_s2}
	\end{center}
\begin{center}
					\includegraphics[width=.75\textwidth]{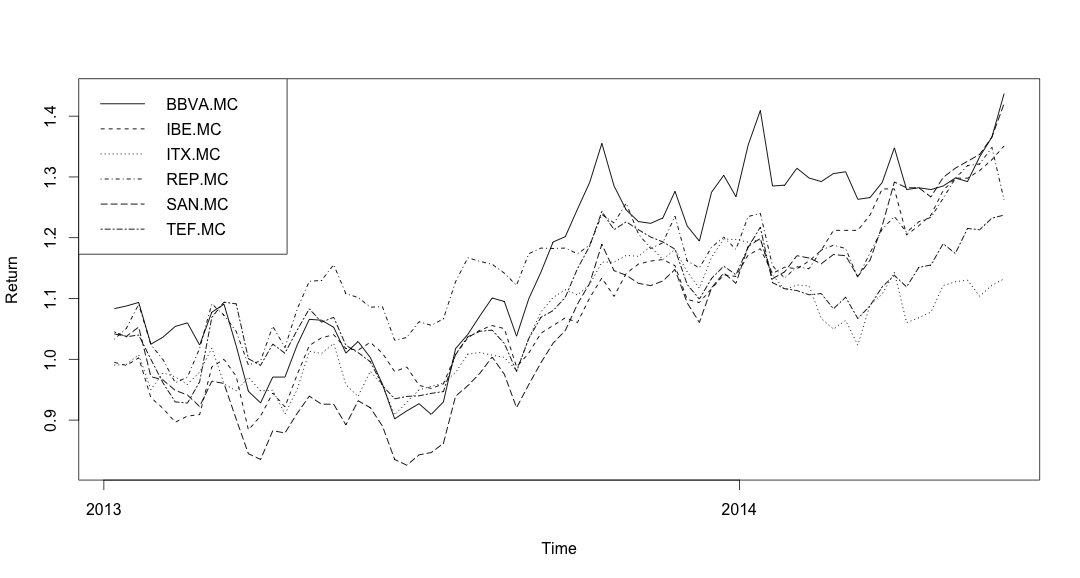}
					\caption{Case 1. Returns  for out-sample period,  from 2013-01-01 to 2014-06-06.}
			\label{fig_s1}
	\end{center}\end{figure}

Table \ref{table1}  shows some descriptive statistics for the set of assets considered: return (average weekly returns), risk (standard deviation of weekly returns), and the Minimum and Maximum return. The first row- block corresponds to the period from 2009 to 2012 (in-sample or estimation period) and the second row-block to the period from 2013 to 2014 (out-sample or test period); being the las row-block the aggregate data from both periods. Figures \ref{fig_s1}, \ref{fig_s2} and \ref{fig_s3} show the accumulated weekly returns for the two periods considered and for the entire period.

\begin{table}[htp]
\caption{Case 1. Weekly descriptive statistics returns for 6 EUROSTOCK assets.  }
\begin{center}
\begin{tabular}{lrrrrrr} \hline
 & BBVA.MC &IBE.MC &ITX.MC & REP.MC & SAN.MC & TEF.MC \\ \hline
& \multicolumn{6}{l}{In-sample (Estimation)  Period (from 2009-01-01 to 2012-12-31)} \\ 
Return & 0.0026 & 0.00044 & 0.0092  & 0.0022  & 0.0034 & 0.00018 \\
Risk   & 0.0623 & 0.04268 & 0.0381  & 0.0455  & 0.0587 & 0.03400 \\  
Minimum & -0.1916 & -0.1483 & -0.1558 & -0.1496 & -0.1760 & -0.0977 \\ 
Maximum & 0.1838 & 0.1385 & 0.1292 & 0.1249 & 0.1916 & 0.1139\\ \hline 
&\multicolumn{6}{l}{Out-sample (Test) Period (from 2013-01-07 to 2014-06-02)} \\ 
Return & 0.0056 & 0.0045 & 0.0021 & 0.0036  & 0.0053 & 0.0034 \\
Risk   & 0.0377 & 0.0287 & 0.0297  & 0.0316  & 0.0335 & 0.0316 \\  
Minimum & -0.0882 & -0.0909 &-0.0723& -0.0697 & -0.0773 & -0.0826  \\ 
Maximum & 0.0943 & 0.0861 & 0.0669 & 0.0719 & 0.0901 & 0.1102\\ \hline 
&\multicolumn{6}{l}{Total period (from 2009-01-01 to 2014-06-02)} \\ 
Return & 0.0035 & 0.0014 & 0.0069  & 0.0028  & 0.0040 & 0.0011 \\
Risk   & 0.0567 & 0.0394 & 0.0365  & 0.0422  & 0.0530 & 0.0333 \\  
Minimum & -0.1916 & -0.1483 & -0.1558 & -0.1496 & -0.1760 & -0.0977 \\ 
Maximum & 0.1838 & 0.1385 & 0.1292 & 0.1249 & 0.1916 & 0.1139\\ \hline 
\end{tabular}
\end{center}
\label{table1}
\end{table}%

In order to evaluate the performance of our solution, the Maximum Cross Efficiency Sharpe Ratio (MCESR), we compare it with the global minimum variance (GMV), tangent (TP) and Maximum Sharpe Ratio (MSR) portfolios. Table \ref{table2}  shows the different solutions evaluated in the in-sample period for the four portfolios considered. The risk-free asset in the interval (0, 0.003) was considered to evaluate the MCESR portfolio, and we chosse the upper limit of the interval considered by the risk free to evaluate the Maximum Sharpe Ration portfolio. Note that the optimal MCESR is obtained when $r_f$ is 0.001773.

Table \ref{table3} shows the reaching value for each out-sample portfolio, i.e., in the test period. Note that we divided the out-sample period in three sub-periods of 25 weeks each in order to evaluate the evolution of the four allocations. In the first 25 weeks, all the portfolios decrease in value, being the minimum variance portfolio (GMV) the best hold. Returns increase  in the next 25 weeks and in this case the portfolio with the higher volatility (MSR) obtains a better performance. Finally, for the entire period out-sample, the GMV portfolio is the only one that provides benefits, and the worst benefit is obtained for the MSR portfolio where the losses outweigh the investment. Note that this last situation is possible because short sales are allowed.

\begin{table}[htp]
\caption{Case 1. In-sample results for different portfolio solutions.}
\begin{center}
\begin{tabular}{lrrrrrr} \hline
\hline
& \multicolumn{6}{l}{GMV -- Global Minimum Variance Portfolio} \\ 
\cline{2-6} & \multicolumn{3}{l}{Expected return =  0.00338} & \multicolumn{3}{l}{Risk = 0.02762} \\ 
\hline
& \multicolumn{6}{l}{TP -- Tangent Portfolio} \\ 
\cline{2-6} & \multicolumn{3}{l}{Expected return =  0.01612} & \multicolumn{3}{l}{Risk = 0.06025} \\ 
\hline
& \multicolumn{6}{l}{MSR -- Maximum Sharpe Ratio Portfolio ($r_f=0.003$)} \\ 
\cline{2-6} & \multicolumn{3}{l}{Expected return =  0.1149} & \multicolumn{3}{l}{Risk = 0.4699} \\ 
\hline
& \multicolumn{6}{l}{MCESR -- Maximum Cross-Efficiency Sharpe Ratio Portfolio ($r_f=0.001773$) } \\ 
\cline{2-6} & \multicolumn{3}{l}{Expected return =  0.02940} & \multicolumn{3}{l}{Risk = 0.1129} \\ 
\hline
\end{tabular}
\end{center}
\label{table2}
\end{table}%

\begin{table}[htp]
\caption{Case 1. Change in portfolio value  for the out-sample  period, from    2014-01-07 to 2014-06-02.}
\begin{center}
\begin{tabular}{lrrr} \hline
Portfolio &  First 25 Weeks  & 50 weeks  & 75 weeks    \\ \hline \hline
GMV       &    -6.1\%    &  5.9\% &  15.1\% \\ \hline
TP &   -11.1\%  &  8.6\%  & -0.3 \% \\ \hline 
 MSR ($r_f=0.003$) &  -49.3\% &  29.0\%  & -119.8\% \\ \hline
 MCESR  ($r_f=0.001773$)  &         -16.2\%   &    11.3\%  & -16.4\%  \\  \hline \hline 
\end{tabular}
\end{center}
\label{table3}
\end{table}%

\begin{figure}[thb]
\begin{center}
					\includegraphics[width=.75\textwidth]{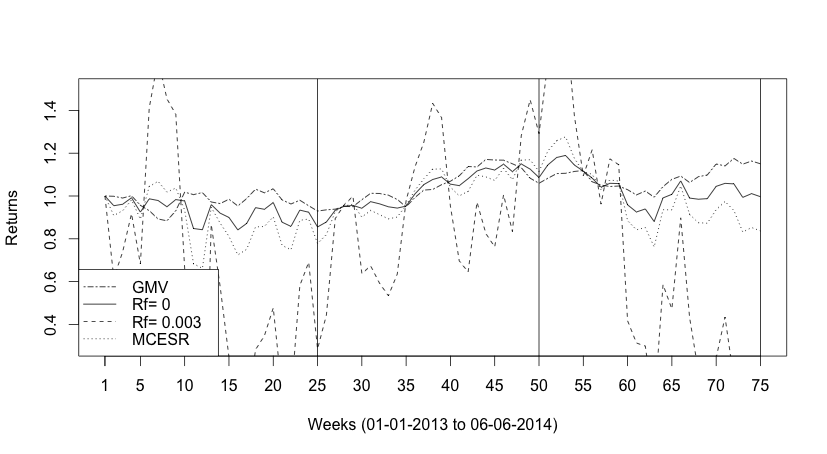}
					\caption{Case 1. Out-sample expected returns from 2013-01-01 to 2013-12-31.}
			\label{fig_t1}
	\end{center}
\end{figure}

Figure  \ref{fig_t1} shows the out-sample performance for the four strategies considered. We see the high volatility associated with the MSR ($r_f = 0.003$) portfolio. Note that although the MCESR portfolio is worse than GMV and TP portfolios, the MCESR provides greater benefits in good times and contains the losses in the bad ones.

\begin{figure}[thb]
\begin{center}
					\includegraphics[width=.75\textwidth]{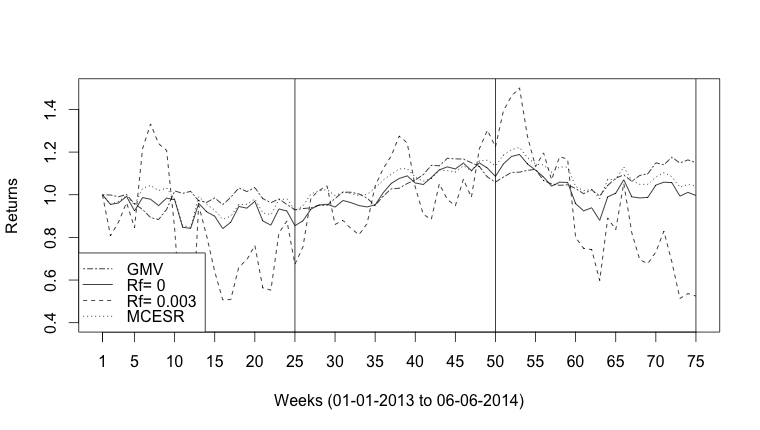}
					\caption{Case 1. Out-sample expected returns  from 2013-01-01 to 2013-12-31.}
			\label{fig_t2}
	\end{center}
\end{figure}

\subsection{Case 2. USA Industry portfolios}

For the second numerical example, we selected ten industry portfolios from the USA market. In the same way as in the case above, we considered two time periods, a first time period to estimate (in-sample period) and a second period (out-sample) to evaluate the performance of each strategy. The data source and the number of observations are shown in Table \ref{table5}. In Table \ref{table6} we keep the same observations as in Table \ref{table1}.

\begin{table}[thb]
\caption{Dataset description.}
\begin{center}
\begin{tabular}{lrrr} \hline
 Case 2 & N & In-sample (estimation) period  & Out-sample (test) period \\   \hline
 Ten industry portfolios.  & 10 &Jan 1963 to Dec 2012 & Jan 2013 to Jul 2014\\
  Monthly data                        &       & (600 observations)    &  (19 observations) \\ \hline
\hline \multicolumn{4}{l}{Source: Ken French's Web Site.} 
\end{tabular}
\end{center}
\label{table5}
\end{table}%

\begin{table}[htp]\small
\caption{Case 2. Monthly  descriptive statistics returns for 10 industry portfolios}
\begin{center}
\begin{tabular}{lrrrrrrrrrr} \hline
 & NoDur &Durbl &Manuf & Enrgy & HiTec & Telcm&Shops&Hlth&Utils&Other \\ \hline
 &\multicolumn{9}{l}{In-sample (estimation) Period (from Jan 1963 to Dec 2012 )} \\ 
Return & 1.1 & 0.86 & 0.97  & 1.1  & 0.96 & 0.86 & 1.0 & 1.1 & 0.83 & 0.92 \\
Risk   & 4.3 & 6.36 & 4.98  & 5.4  & 6.59 & 4.67 & 5.2 & 4.9 & 4.03 & 5.35  \\  
Minimum & -21.03 & -32.63 & -27.33 & -18.33 & -26.01 & -16.22 & -28.25 & -20.46 & -12.65 & -23.6  \\ 
Maximum & 18.88 & 42.62 & 17.51 & 24.56 & 20.75 & 21.34 & 25.85 & 29.52 & 18.84 & 20.22 \\ \hline 
&\multicolumn{9}{l}{Out-sample (test) Period (Jan 2013 to Jul 2014 )} \\ 
Return & 1.4 & 2.3 & 1.7  & 1.7  & 2.0 & 1.9 & 1.5 & 2.4 & 1.4 & 1.9 \\
Risk   & 3.3 & 4.0 & 3.2  & 3.7  & 2.5 & 2.8 & 3.5 & 3.4 & 3.8 & 3.2  \\  
Minimum & -5.71 & -4.6 & -4.33 & -6.97 & -2.84 & -3.94 & -6.65 & -3.67 & -6.96 & -4.37  \\ 
Maximum & 5.21 & 9.9 & 6.03 & 7.71 & 5.97 & 5.62 & 5.97 & 8.11 & 5.51 & 6.87 \\ \hline 
&\multicolumn{9}{l}{Total Period (Jan 1963 to Jul 2014 )} \\ 
Return & 1.1 & 0.9 & 0.99  & 1.1  & 1.0 & 0.89 & 1.0 & 1.1 & 0.85 & 0.95 \\
Risk   & 4.3 & 6.3 & 4.93  & 5.4  & 6.5 & 4.63 & 5.2 & 4.9 & 4.02 & 5.30  \\  
Minimum & -21.03 & -32.63 & -27.33 & -18.33 & -26.01 & -16.22 & -28.25 & -20.46 & -12.65 & -23.6  \\ 
Maximum & 18.88 & 42.62 & 17.51 & 24.56 & 20.75 & 21.34 & 25.85 & 29.52 & 18.84 & 20.22 \\ \hline 
\end{tabular}
\end{center}
\label{table6}
\end{table}%

\begin{figure}[thb]
\begin{center}
					\includegraphics[width=.75\textwidth]{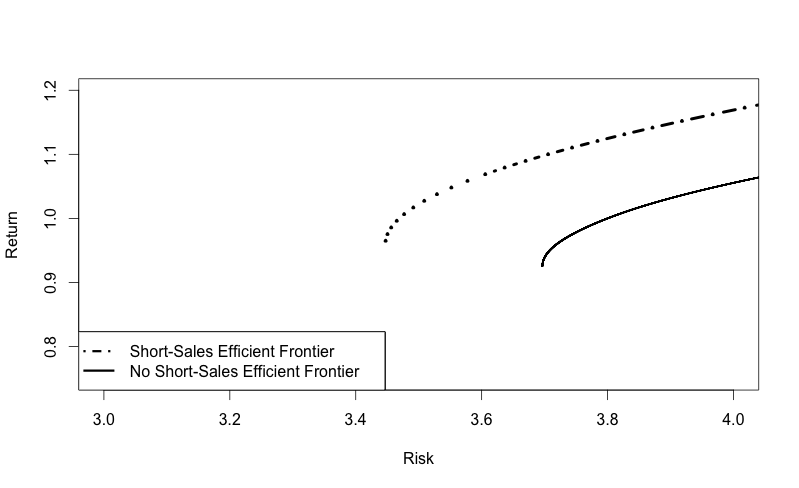}
					\caption{Case 2.  Efficient Frontiers for ten industry portfolios.}
			\label{fig_fronteras}
	\end{center}
\end{figure}

Table \ref{table7}  shows the different portfolios in the in-sample period, and for each portfolio we  report their allocation. The risk-free asset in the interval (0,0.9) was considered in order to calculate the MCESR portfolio, wich leads to obtaining the optimal MCESR $r_f$ is 0.57103. If short sales are not allowed, the optimal MCESR is obtained when $r_f$ is 0.576. See Table \ref{table8} for the same results as in  Table \ref{table7} when short sales are not allowed. Figure \ref{fig_fronteras} shows the efficient frontier of Malkowitz for both cases, with and without short sales.

\begin{table}[htp] \footnotesize
\caption{Case 2. In-sample results for different portfolio solutions.  10 industry portfolios. Short-Sales}
\begin{center}
\begin{tabular}{lrrrrrrrrrr} \hline
& NoDur &Durbl &Manuf & Enrgy & HiTec & Telcm&Shops&Hlth&Utils&Other \\ \hline
& \multicolumn{10}{l}{GMV -- Global Minimum Variance  portfolio} \\ 
\cline{2-6} & \multicolumn{3}{l}{Expected return =  0.96} & \multicolumn{3}{l}{Risk = 3.45} \\ 
\cline{2-11}  Allocation   & 0.29 & 0.00 & 0.09  & 0.11  & 0.02& 0.26 & 0.08 & 0.15 & 0.45 & -0.44 \\   \hline 
\hline
& \multicolumn{10}{l}{TP -- Tangent portfolio} \\ 
\cline{2-6} & \multicolumn{3}{l}{Expected return =  1.09} & \multicolumn{3}{l}{Risk = 3.68} \\ 
\cline{2-11}  Allocation   & -0.70 &-0.05& -0.05&  0.30&  0.03&  0.18&  0.14 & 0.17&  0.17& -0.58 \\   \hline 
\hline
& \multicolumn{10}{l}{MSR -- Maximum Sharpe Ratio portfolio ($r_f=0.9$)} \\ 
\cline{2-6} & \multicolumn{3}{l}{Expected return =  3.08} & \multicolumn{3}{l}{Risk = 20.84} \\ 
\cline{2-11}  Allocation   & 6.85 &-0.77& -2.13&  3.19&  0.24& -1.04&  1.01&  0.53 &-4.15 &-2.72  \\  \hline 
\hline
& \multicolumn{10}{l}{MCESR -- Maximium Cross-Efficincy Sharpe Ratio portfolio  ($r_f=0.57103$) } \\ 
\cline{2-6} & \multicolumn{3}{l}{Expected return =  1.29} & \multicolumn{3}{l}{Risk = 4.67} \\ 
\cline{2-11}  Allocation   & 1.29 &-0.12& -0.25&  0.58&  0.05&  0.06&  0.22&  0.21& -0.26& -0.79 \\  \hline 
\hline
\end{tabular}
\end{center}
\label{table7}
\end{table}%

\begin{table}[htp] \footnotesize
\caption{Case 2. In-sample results for different portfolio solutions.  10 industry portfolios.  No Short-Sales}
\begin{center}
\begin{tabular}{lrrrrrrrrrr} \hline
 & NoDur &Durbl &Manuf & Enrgy & HiTec & Telcm&Shops&Hlth&Utils&Other \\ \hline
& \multicolumn{10}{l}{GMV --  Gloabal Minimum Variance  portfolio} \\ 
\cline{2-6} & \multicolumn{3}{l}{Expected return = 0.93 } & \multicolumn{3}{l}{Risk = 3.70} \\ 
\cline{2-11}  Allocation   &0.30 & 0.00&  0.03&  0.00&  0.00&  0.14&  0.00&  0.06&  0.47&  0.00\\   \hline 
\hline
& \multicolumn{10}{l}{TP -- Tangent portfolio} \\ 
\cline{2-6} & \multicolumn{3}{l}{Expected return =  1.03} & \multicolumn{3}{l}{Risk = 3.90} \\ 
\cline{2-11}  Allocation   & 0.53 & 0.00&  0.02&  0.07&  0.00&  0.00&  0.03 & 0.19&  0.15&  0.00  \\   \hline 
\hline
& \multicolumn{10}{l}{MSR -- Maximum Sharpe Ratio portfolio ($r_f=0.9$)} \\ 
\cline{2-6} & \multicolumn{3}{l}{Expected return = 1.076 } & \multicolumn{3}{l}{Risk =4.14 } \\ 
\cline{2-11}  Allocation   & 0.67 & 0.00&  0.00&  0.23&  0.00&  0.00&  0.00&  0.10&  0.00&  0.00  \\  \hline 
\hline
& \multicolumn{10}{l}{MCESR -- Maximum Cross-Efficiency Sharpe Ratio portfolio ($r_f=0.576$)} \\ 
\cline{2-6} & \multicolumn{3}{l}{Expected return =  1.07} & \multicolumn{3}{l}{Risk = 4.09 } \\ 
\cline{2-11}  Allocation   &  0.63&  0.00&  0.00&  0.14&  0.00&  0.00&  0.00&  0.22&  0.00&  0.00  \\  \hline \hline
\end{tabular}
\end{center}
\label{table8}
\end{table}%

\newpage

In order to compare the performance of four strategies, we evaluated them in the out-sample period. The expected returns for each portfolio, with and without short sales, are shown in Table \ref{table9}. Note that if short sales are allowed, the $MSR(r_f = 0.9)$ portfolio originates losses of 14.3\%, while the same strategy portfolio causes a benefit of 33.5\% if short sales are not allowed.

The portfolio with less variation with or without short sales is the GMV portfolio. The MCESR portfolio provides a profit of 25.5\% with short sales, and 36.1\% without short sales, being this last profit the highest value for all portfolios considered in both situations.

\begin{table}[htp]
\caption{Case 2. Change in portfolio value  for the period from   2014-01-07 to 2014-06-02.}
\begin{center}
\begin{tabular}{lrrr} \hline
Portfolio &  Short Sales  & No Short Sales    \\ \hline \hline
GMV       &    32.7\%    &  32.9\% \\ \hline
Tangent Portfolio &   29.8\%  &   35.1\% \\ \hline 
 MSR ($r_f=0.9$) &  -14.3\% &  33.5\% \\ \hline
 MCESR &        25.5\%   &    36.1\%  \\  \hline \hline 
\end{tabular}
\end{center}
\label{table9}
\end{table}%

\begin{figure}[htp]
\begin{center}
					\includegraphics[width=.75\textwidth]{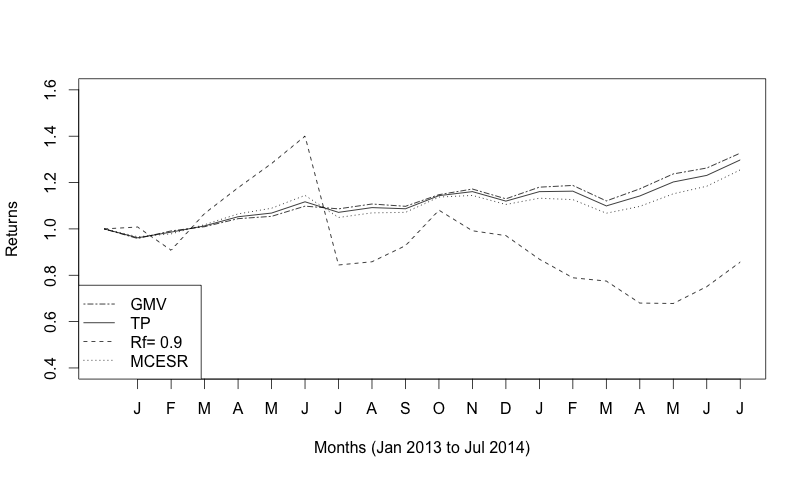}
					\caption{Case 2.  Returns from 2014-01-07 to 2014-06-02. With Short Sales}
			\label{fig_t3}
	\end{center}

\begin{center}
					\includegraphics[width=.75\textwidth]{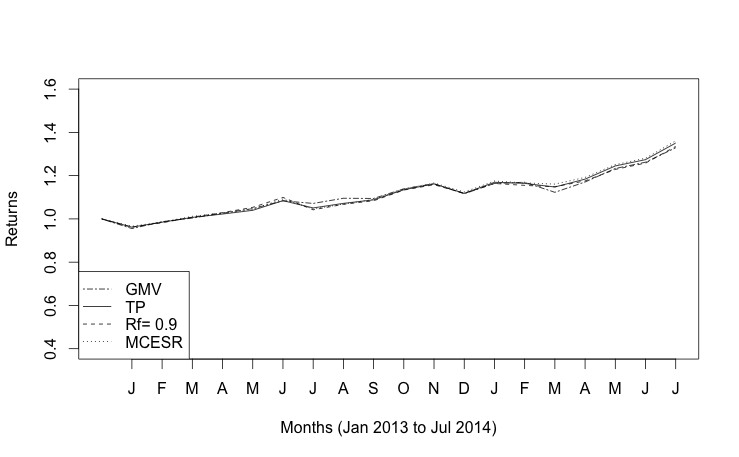}
					\caption{Case 2.  Returns from 2014-01-07 to 2014-06-02. Without Short Sales}
			\label{fig_t4}
	\end{center}
\end{figure}

Figures \ref{fig_t3} and  \ref{fig_t4} show the out-sample performance for the four strategies considered. Note the high volatility of the MSR portfolio when short sales are allowed in front to the homogeneity of the rest. If short sales are not allowed, the four portfolios present practically the same curve, although in this case the MCESR provides the best performance.

\newpage

\section{Conclusions}

This paper proposes a new portfolio selection strategy based on a cross-efficiency evaluation. We compare the new allocation with the classic global minimum and tangent portfolios through a numerical study. The results  show that our allocation is comparable with the others in terms of performance in the out-sample period.

We have derived an explicit expression for the MCESR portfolio when short sales are allowed, and proposed procedures to obtain it when short sales are not allowed. We have also found a relationship between the slopes of the three portfolios considered and that between the MCESR portfolio with the expected returns of the GMV and TP portfolios.

For future research, we plan to apply this new portfolio solution (MCESR) to a large testbed in order to investigate their advantages over the rest.

 \section*{Appendix}
 \proof{Proof of Proposition 2.} {\em The efficient portfolio $i=(\sigma^*_{MSR_i}, r^*_{MSR_i})$ that maximizes the cross-efficiency $CE_i$, in the interval  $[r_{1},r_{2}]$, is reached when 
  \begin{equation}
r_i^* =  r_{GMV}^*+\sigma_{GMV}^* \frac{  \displaystyle \frac{ \displaystyle r^*_{MSR_{2}} - r_{2}}{\sigma^*_{MSR_{2}}}  - \frac{r^*_{MSR_{1}} - r_{1}}{\sigma^*_{MSR_{1}}}  }{ \ln \left ( \displaystyle \frac{ 
\displaystyle \frac{   \displaystyle r^*_{MSR_{2}} - r_{2}}{ \displaystyle \sigma^*_{MSR_{2}}}  - \frac{ r^*_{GMV} - r_{2}}{\sigma^*_{GMV}  }}{ 
\displaystyle    \frac{ \displaystyle r^*_{MSR_{1}} - r_{1}}{\sigma^*_{MSR_{1}}}  -\frac{ \displaystyle r^*_{GMV} - r_{{1}}}{\sigma^*_{GMV} }      } \right )}
 \nonumber
\end{equation}
}

The cross efficiency of portfolio $i$, $CE_i$, depends of the risk-free rate, $r_i$, associated with the portfolio $i$. We can considered $CE_i$ as a function of $r_i$, for $r_i\in [r_{\min},r_{\max}]$. We can write $CE_i(r_i)$ as follows

$$
CE_i(r_i)=\displaystyle \frac{\displaystyle r^*_{MSR_i}}{\sigma^*_{MSR_i}} \int_{r_{\min}}^{r_{\max}}\frac{ \displaystyle \sigma^*_{MSR}}{r^*_{MSR}-r_f} \, \text{d}r_{f}  - \frac{1}{\displaystyle \sigma^*_{MSR_i}} \int_{r_{\min}}^{r_{\max}}\frac{ \displaystyle\sigma^*_{MSR} \, \,  r_f}{ \displaystyle r^*_{MSR}-r} \, \text{d}r_{f} $$

From expressions (\ref{sol_GMV}), (\ref{sol1_MSR}) and (\ref{sol2_MSR}), using notation of  (\ref{abc}), we can derivate the following identities for the expected return and variance of GMV and MSR portfolios:
\begin{align}
&r_{GMV}^*=\frac{c}{b}, \quad \sigma_{GMV}^*=\frac{1}{\sqrt{b}},\quad \frac{r_{GMV}^*-r_f}{\sigma^{*2}_{GMV}}=c-b\,\, r_f, \quad \frac{r_{MSR}^*}{\sigma_{MSR}^*}= \frac{a-c \,\, r_f}{\sqrt{a-2c \,\, r_f+b \,\, r_f^2 }}, \nonumber \\
 &\frac{1}{\sigma_{MSR}^*}= \frac{c-b\,\, r_f}{\sqrt{a-2c\,\, r_f+b\,\, r_f^2 }}, \quad  \text{and} \quad \frac{\sigma_{MSR}^*}{r_{MSR}^*-r_f}= \frac{1}{\sqrt{a-2c\,\, r_f+b\,\, r_f^2 }}, \label{identities} 
 \end{align}
and write the the cross efficiency, $CE_i(r_i)$, in terms of variable $r_i$.
$$
CE_i(r_i)=\displaystyle \frac{\displaystyle a-c\,\, r_i}{\sqrt{a-2c\,\, r_i+b\,\, r_i^2}} I_1   - \frac{c-b\,\, r_i}{\sqrt{a-2c\,\, r_i+b \,\, r_i^2}} I_2$$
where 
$$
I_1=\frac{1}{r_{\max}-r_{\min}}\int_{r_{\min}}^{r_{\max}}\frac{ \displaystyle \text{d}r_{f} }{\sqrt{a-2c \,\, r_f+b \,\, r_f^2}} 
\quad \text{and} \quad 
I_2=\frac{1}{r_{\max}-r_{\min}} \int_{r_{\min}}^{r_{\max}}\frac{r_f\,\, \, \text{d}r_{f} }{\sqrt{a-2c\,\, r_f+b \,\, r_f^2}}
$$
The function $CE_i(R_i)$ has first derivate 

\begin{align}
CE'_i(r_i)=& \frac{-c \sqrt{a-2cr_i+br_i^2} -(a-cr_i)(br-c)/\sqrt{a-2cr_i+br_i^2}}{\left (\sqrt{a-2cr_i+br_i^2}\right )^2}I_1- \nonumber\\
                 &-\frac{-b\sqrt{a-2cr_i+br_i^2}-(c-br_i)(br_i-c)/\sqrt{a-2cr_i+br_i^2}}{\left (\sqrt{a-2cr_i+br_i^2}\right )^2}I_2 \nonumber \\
                 =& \frac{(c^2r_i-abr_i)I_1+(ba-c^2)I_2}{\left (\sqrt{a-2cr_i+br_i^2}\right )^3}. \nonumber
\end{align}

It is left to show that $CE'_i(r_i)=0$ for $r_i=I_2/I_1$, therefore, $r_i=I_2/I_1$ is a point with slope zero, and it is a candidate to a maximum in the interval $[r_{\min},r_{\max}]$. The second derivate of the function $CE_i(r_i)$ is given by the following  expression
\begin{align}
CE''_i(r_i)=& \frac{(c^2-ab)I_1\left (\sqrt{a-2cr_i+br_i^2}\right )^3}{\left (\sqrt{a-2cr_i+br_i^2}\right )^6}
		 - \frac{3\left ((ba-c^2)I_2+(c^2-ab)I_1r_i \right )(br_i-c)}{\left (\sqrt{a-2cr_i+br_i^2}\right )^5}  \label{second.d}
\end{align}
and  the second term of (\ref{second.d}) is zero at $r_i=I_2/I_1$, and 
$$
CE''(I_2/I_1)= \frac{(c^2-ab)I_1}{(a-2c \, I_2/I_1+b \, I_2^2/I_1^2)}
$$
Since $\Sigma^{-1}$ is positive definite matrix, then  $(\mu-r)^T\Sigma^{-1}(\mu-r)= a-2c r +b r^2 > 0$, with  discriminant  $4(c^2-ab) < 0$, then   the second derivate at $r_i=I_2/I_1$, $CE''(I_2/I_1)$, is less to 0.

Next, we show the expression of $I_2/I_1$.
\begin{align}
(r_{\max}-r_{\min}) I_1&=\int_{r_{\min}}^{r_{\max}}\frac{ \displaystyle \text{d}r_{f} }{\sqrt{a-2cr_f+br_f^2}} =\left [   \frac{1}{\sqrt{b}}  \ln \left (  \sqrt{b} \sqrt{a-2cr_f+br_f^2} + b r_f -c  \right )\right ]_{r_{\min}}^{r_{\max}} \nonumber \\
(r_{\max}-r_{\min})  I_2&= \int_{r_{\min}}^{r_{\max}}\frac{r_f\, \text{d}r_{f} }{\sqrt{a-2cr_f+br_f^2}}=\nonumber \\
         &=  \left [   \frac{c}{\sqrt{b}^3}  \ln \left (  \sqrt{b} \sqrt{a-2cr_f+br_f^2} + b r_f -c  \right )  + \frac{1}{b}\sqrt{a-2cr_f+br_f^2}   \right ]_{r_{\min}}^{r_{\max}} \nonumber
\end{align}

Now, we can write the above expression using the identities of (\ref{identities}) as follows:
 
\begin{align}
(r_{\max}-r_{\min}) I_1&=  \sigma_{GMV}^* \ln  \left ( \displaystyle \frac{ 
\displaystyle \frac{   \displaystyle r^*_{MSR_{2}} - r_{2}}{ \displaystyle \sigma^*_{MSR_{2}}}  - \frac{ r^*_{GMV} - r_{2}}{\sigma^*_{GMV}  }}{ 
\displaystyle    \frac{ \displaystyle r^*_{MSR_{1}} - r_{1}}{\sigma^*_{MSR_{1}}}  -\frac{ \displaystyle r^*_{GMV} - r_{{1}}}{\sigma^*_{GMV} }      } \right  )   \nonumber \\
(r_{\max}-r_{\min})  I_2&= r_{GMV}^* \sigma_{GMV}^* \ln  \left ( \displaystyle \frac{ 
\displaystyle \frac{   \displaystyle r^*_{MSR_{2}} - r_{2}}{ \displaystyle \sigma^*_{MSR_{2}}}  - \frac{ r^*_{GMV} - r_{2}}{\sigma^*_{GMV}  }}{ 
\displaystyle    \frac{ \displaystyle r^*_{MSR_{1}} - r_{1}}{\sigma^*_{MSR_{1}}}  -\frac{ \displaystyle r^*_{GMV} - r_{{1}}}{\sigma^*_{GMV} }      } \right  ) + \nonumber \\ 
&+ \sigma_{GMV}^{*2} \left( \frac{r_{MSR_{2}}^*-r_{2}}{\sigma^*_{{MSR}_2}} -  \frac{r_{MSR_{1}}^*-r_{1}}{\sigma^*_{{MSR}_1}}  \right ) \nonumber
\end{align}
and, finally we can write the maximum $r_i$ as 
  \begin{equation}
r_i^* =  r_{GMV}^*+\sigma_{GMV}^* \frac{  \displaystyle \frac{ \displaystyle r^*_{MSR_{2}} - r_{2}}{\sigma^*_{MSR_{2}}}  - \frac{r^*_{MSR_{1}} - r_{1}}{\sigma^*_{MSR_{1}}}  }{ \ln \left ( \displaystyle \frac{ 
\displaystyle \frac{   \displaystyle r^*_{MSR_{2}} - r_{2}}{ \displaystyle \sigma^*_{MSR_{2}}}  - \frac{ r^*_{GMV} - r_{2}}{\sigma^*_{GMV}  }}{ 
\displaystyle    \frac{ \displaystyle r^*_{MSR_{1}} - r_{1}}{\sigma^*_{MSR_{1}}}  -\frac{ \displaystyle r^*_{GMV} - r_{{1}}}{\sigma^*_{GMV} }      } \right )}
\label{r.f}
\end{equation}

 \Halmos
\endproof 

  \proof{Proof of Proposition 3.} {\em
  Exist a Pythagorean  relationship  between the slopes of the Tangent  and Global Minimum  portfolios and the slope of the asymptote of $W_{\rho}$. 
 \begin{equation}
m_{TP}^2= m_{ah}^2+m_{GMV}^2 
\end{equation}
}

From expressions  (\ref{abc}), we can derivate the following identities for the  $m_{TP}$, $m_{ah}$ and $m_{GMV}$ slopes: 

\begin{align}
&m_{TP}=\sqrt{a}, \qquad m_{ah}= \sqrt{\frac{ab-c^2}{b}} , \quad \text{and} \quad m_{GMV}=\frac{c}{\sqrt{b}}.   \label{mmm}
\end{align}
and now, we can derivate the relationship $m_{TP}^2= m_{ah}^2+m_{GMV}^2$, 

\begin{align}
 m_{ah}^2+m_{GMV}^2=  \frac{ab-c^2}{b} + \frac{c^2}{b}  = a=  m_{TP}^2 \nonumber 
 \end{align} 
 \Halmos
\endproof

  \proof{Proof of Corollary 2.} {\em 
  The maximum cross-efficiency (MCESR) portfolio  in $[0,r_{GMV}^*]$ depends only  of Minimal Global Variance and Tangent portfolios. 
   }
   
   \begin{equation}
r_i^* =  r^*_{GMV}\left ( 1- 
\frac{\displaystyle   \sqrt{\frac{r^*_{TP}}{r^*_{GMV} }}-   \sqrt{\frac{r^*_{TP}}{r^*_{GMV} }-1}     }
{ \ln \left ( \displaystyle  \sqrt{\frac{r^*_{TP}}{r^*_{GMV} } -1}  \right )   - \ln \left (\displaystyle  \sqrt{\frac{r^*_{TP}}{r^*_{GMV} }} -1 \right )   }
\right ) \label{r4p.f}
\end{equation}

From (\ref{identities}) and (\ref{mmm}), we can derivate the following expressions:
\begin{align}
&\frac{m_{TP}^2}{m_{GMV}^2}=\frac{a}{c^2/\sqrt{b}^2}= \frac{a/c}{c/b}= \frac{r^*_{TP}}{r^*_{GMV}}, \, \text{then} \quad \frac{m_{TP}}{m_{GMV}}=\sqrt{\frac{r^*_{TP}}{r^*_{GMV}}}  \nonumber \\
&\frac{m_{ah}^2}{m_{GMV}^2}= \frac{ \frac{ab-c^2}{b}}{c^2/b} = \frac{ab-c^2}{c^2}=\frac{ab}{c^2}-1= \frac{r^*_{TP}}{r^*_{GMV}} -1,  \, \text{then} \quad \frac{m_{ah}}{m_{GMV}}=\sqrt{\frac{r^*_{TP}}{r^*_{GMV}}-1} \nonumber \\ 
&\frac{m_{TP}^2}{m_{ah}^2}=\frac{a}{\frac{ab-c^2}{b}}=\frac{a/c}{a/c-c/b} =\frac{r^*_{TP}}{r^*_{TP}-r^*_{GMV}},  \, \text{then} \quad \frac{m_{TP}}{m_{ah}}= \sqrt{\frac{r^*_{TP}}{r^*_{TP}-r^*_{GMV}}} \nonumber 
\end{align}
From  the expression (\ref{r3.f}),   it is left to show that (\ref{r4p.f}) is true.
 \Halmos
\endproof 

\newpage

\end{document}